\documentclass{mem}
\usepackage{natbib}\usepackage{txfonts}\usepackage{balance}
\usepackage{graphicx}
\usepackage[a4paper,breaklinks,dvipdfm]{hyperref}
\idline{75}{282}
\begin{document}
\def\teff{$T\rm_{eff }$}
\def\kms{$\mathrm {km s}^{-1}$}

\title{
A Near-Infrared Spectroscopic Study of Young Field Ultracool Dwarfs: Additional Analysis
}

   \subtitle{}

\author{
K.~N.~Allers\inst{1} 
\and Michael~C.~Liu\inst{2}
          }

 % \offprints{K.~N.~Allers}

\institute{
Department of Physics and Astronomy, Bucknell University, Lewisburg, PA 17837, USA; \email{k.allers@bucknell.edu}
\and
Institute for Astronomy, University of Hawaii, 
2680 Woodlawn Drive, Honolulu, HI 96822, USA
}

\authorrunning{Allers \& Liu}

\titlerunning{Near-IR Spectra of Young UCDs}

\abstract{We present additional analysis of the classification system presented in \citet{allers13}.  We refer the reader to \citet{allers13} for a detailed discussion of our near-IR spectral type and gravity classification system.  Here, we address questions and comments from participants of the Brown Dwarfs Come of Age meeting. In particular, we examine the effects of binarity and metallicity on our classification system.  We also present our classification of Pleiades brown dwarfs using published spectra.  Lastly, we determine SpTs and calculate gravity-sensitive indices for the BT-Settl atmospheric models and compare them to observations.

}
\maketitle{}

\section{Introduction}

In \citet{allers13}, hereinafter A13, we present a method for classifying the spectral types (SpTs) and surface gravities of ultracool dwarfs.  The SpT classification utilizes both visual comparison to field standards and SpT-sensitive indices.  A13 also present new gravity-sensitive indices which measure FeH, VO, alkali line and $H$-band continuum features.  Using gravity-sensitive indices and line EWs, A13 propose three near-IR gravity classes:  {\sc fld-g} for objects with normal field dwarf gravities, {\sc vl-g} for objects with strong spectral signatures of youth (ages~$\sim$10--30~Myr), \& {\sc int-g} for objects with intermediate spectral signatures of youth (ages~$\sim$50--200~Myr).

\section{Binarity}

Unresolved binarity can cause peculiarities in the near-IR spectra of brown dwarfs, which are apparent even at low spectral resolution (R $\approx$ 100).  Spectral peculiarity has been used to identify candidate brown dwarf binaries \citep[e.g.,][]{burgasser10}.  Young, low-gravity objects also show signs of spectral peculiarity, which raises two interesting questions: 1) could the spectral peculiarities we attribute to low-gravity be mimicked by unresolved binarity of normal field dwarfs?  and 2) to what extent could binarity affect our classification of young, low-gravity, ultracool dwarfs?

To test if the spectra of unresolved field dwarf binaries could show evidence of youth in our indices, we combined the spectra of the field dwarf near-IR standards from \citet{kirkpatrick10} to create artificial binaries. We first scaled the spectra of the field standards using SpT - M$_J$ relations from \citet{dupuy12} so that the spectra were in units of absolute flux.  We then co-added two scaled spectra for all possible pairings of standards to  create 434 field composite binary spectra.  Using the method of A13, we determined the near-IR SpTs for each binary and found that we properly classified all of our artificial M4--L6 (the applicable range of the A13 method) field dwarf binaries as having normal {\sc fld-g} gravities.  We conclude that normal field dwarf binaries are unlikely to contaminate spectroscopic samples of young-low gravity objects.

To test the effects of binarity on our classification of low-gravity objects, we created artificial composite binary spectra by combining the low-resolution spectra of young objects in the A13 sample having published parallax values.  Table \ref{plx} lists the particular spectra we used.  
We created the artificial low-gravity binary spectra in a manner similar to that used to create artificial field dwarf binary spectra, except that we scaled each low-gravity spectrum to absolute flux units using published parallaxes and $JHK$ mags.  We then determined the SpTs and gravities of the artificial low-gravity binary spectra using the methods outlined in A13.   The SpTs of the artificial binaries were found to agree with the near-IR SpTs of the primary star to within 1 subtype.  The gravity classifications for 54 of the 55 low-gravity artificial binaries agreed with the gravity classifications of the low-resolution spectra of the primaries.   The only simulated binary whose classification did not agree with its primary was 2M~0032-44 + 2M~0355+11, which we classify as L1 {\sc int-g}. Overall, it appears that binarity does not significantly affect our SpT or gravity classifications.

\begin{table}
\caption{Objects Used for Binary Simulations}
\label{plx}
%\begin{center}
\begin{tabular}{lcrl}
\hline
\\
Object & SpT$^a$ & $M_J$ & Ref\\
\hline
\\
TWA~27A      & M8 {\sc vl-g} & 9.4 & M03 \\
TWA~26        & M9 {\sc vl-g} & 9.5 & W13 \\
TWA~29        & L0 {\sc vl-g} & 10.0 & W13 \\
2M~0608-27 & L0 {\sc vl-g} & 11.1 & F12  \\
2M~0518-27 & L1 {\sc vl-g} & 11.8 & F12 \\
PC~0025+0447 & L0 {\sc int-g} & 11.9 & D02 \\
2M~0032-44 & L0 {\sc vl-g} & 12.6 & F12 \\
2M~0536-19 & L2 {\sc vl-g} & 12.7 & F12 \\
2M~0355+11 & L3 {\sc vl-g} & 14.3 & L13 \\
2M~0501-00 & L3 {\sc vl-g} & 14.3 & F12 \\
2M~0103+19 & L6 {\sc int-g} & 14.5 & F12 \\
\hline
\end{tabular}

$^a$Near-IR spectral types and gravities from A13.

{\bf References:}
D02=\citet{dahn02};
M03=\citet{ducourant08};
F12=\citet{faherty12};
L13=\citet{liu13};
W13=\citet{weinberger13}
%\end{center}
\end{table}

\section{Metallicity} 

\begin{figure}[]
\resizebox{\hsize}{!}{\includegraphics[clip=true]{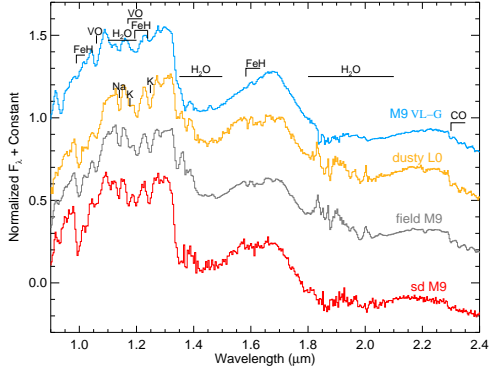}}
\caption{
\footnotesize
Comparison of M9--L0 ultracool dwarfs.  From top to bottom, the spectra are TWA~26 \citep{looper07}, 2M~1331+34 \citep{kirkpatrick10}, LHS~2924 \citep{kirkpatrick10} and 2M~0041+35 \citep{burgasser04}.  Using the system of A13, we classify 2M~0041+35 as L0 {\sc fld-g}.  The 0.98 $\mu$m FeH feature is significantly stronger in the subdwarf spectrum compared to other ultracool dwarf spectra of similar SpT.
}
\label{sdl0}
\end{figure}

In A13, we did not consider the effects of metallicity when determining the SpTs and gravity classifications for our sample.  Our gravity-sensitive indices measure the depths of FeH, alkali line (Na and K) and VO features, which in addition to being gravity dependent, are sensitive to metallicity \citep[e.g.,][]{mann13, kirkpatrick10}.  Figure \ref{sdl0} compares the spectrum of a mildly metal-poor object \citep[2M~0041+35;][]{burgasser04} to the spectra of young, dusty, and normal field ultracool dwarfs of similar optical SpT.  The A13 classification system types this object as an L0 {\sc fld-g}, in good agreement with its optical spectral classification.  

\begin{figure}[]
\resizebox{\hsize}{!}{\includegraphics[clip=true]{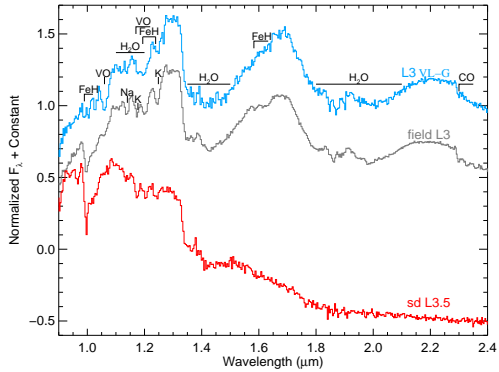}}
\caption{
\footnotesize
Comparison of L3 ultracool dwarfs.  From top to bottom, the spectra are 2M~2208+29 (A13), 2M~1506+13 \citep{burgasser07}, and SDSS~1256-02 \citep{burgasser09}.  Using the system of A13, we classify SDSS~1256-02 as M6 {\sc fld-g}.
}
\label{sdl3}
\end{figure}

Not all subdwarfs are well classified by the A13 system, however.  Figure \ref{sdl3} compares the spectra of low-gravity, normal and subdwarf L3-L3.5 objects.  Although the subdwarf, SDSS~1256-02 \citep{burgasser09}, is classified as {\sc fld-g}, its near-IR SpT is determined to be M6, in stark contrast to its optical type of sdL3.5.  We often determined near-IR SpTs of subdwarfs that are significantly earlier than their published optical SpTs.  Thus, if one suspects a spectrum could be low metallicity, extreme caution should be used when determining near-IR SpTs.

Figure \ref{indices} displays the indices calculated for subdwarf spectra, all of which are classified as {\sc fld-g}.  We note that the A13 study included several ``dusty'' brown dwarfs, whose spectral peculiarities could be due to a metal-rich photosphere \citep{looper08}, all of which were classified as {\sc fld-g}.  Thus, it does not appear that high or low metallicity ultracool dwarfs would be misclassified by A13 as having low gravity.

\section{Pleiades Brown Dwarfs}

\begin{figure*}[t!]
\resizebox{\hsize}{!}{\includegraphics[clip=true]{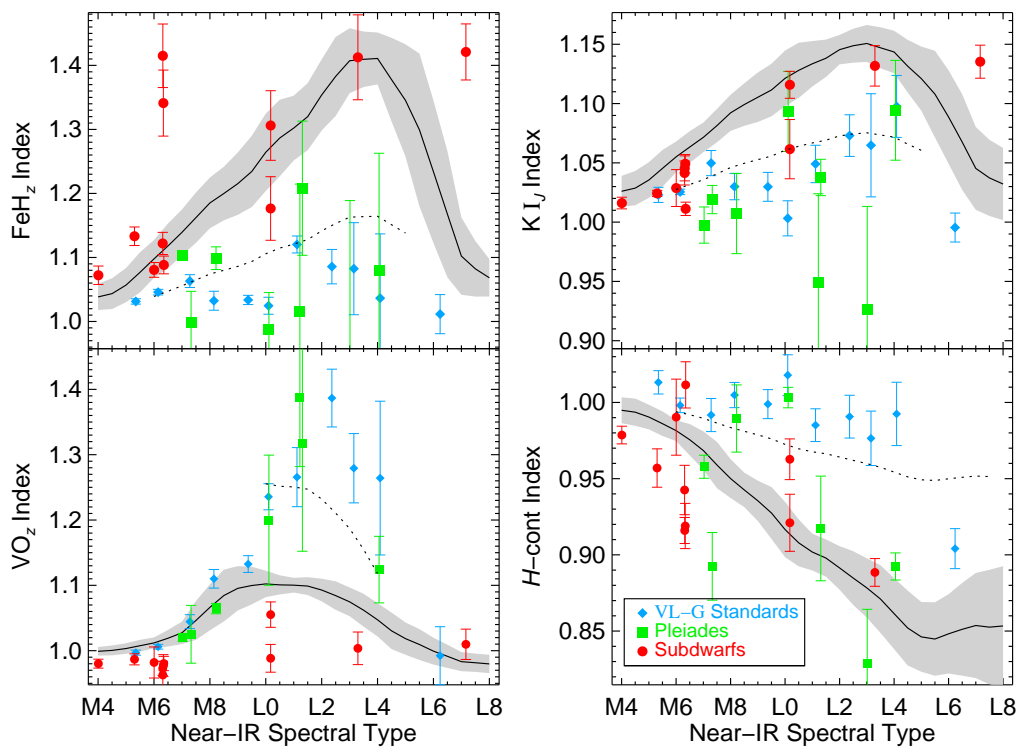}}
\caption{\footnotesize
Gravity-sensitive indices of A13.  Diamond points are for the {\sc vl-g} standards proposed by A13.  Squares are indices calculated for Pleiades brown dwarfs from \citet{bihain10}.  Circles show the indices calculated for subdwarfs with optical SpTs of M7 and later \citep{burgasser04,burgasser06, bowler09, kirkpatrick10} .  All near-IR spectral types are calculated using the method described in A13.
}
\label{indices}
\end{figure*}

In A13, we claim that our classification system can identify low-gravity brown dwarfs with ages $\lesssim$200~Myr.  To test this, we classified spectra for ultracool Pleiades dwarfs from \citet{bihain10}.   We note that many of the spectra in \citet{bihain10} have low S/N ($\lesssim$20) compared to the spectra in the A13 sample.  Table \ref{pleiades} shows the results of our classification.  We calculate SpTs for the objects that are in agreement with the \citet{bihain10} SpTs to within $\pm$1~subtype.  We classify all of the Pleiades objects as having low-gravity (and most as having {\sc vl-g}).  It is interesting to note that among Pleiades spectra of similar SpT, the features indicating youth vary among the objects (as indicated by which features receive scores of ``2'' in Table \ref{pleiades}), with the caveat that some calculated indices have low S/N (Figure \ref{indices}).  This supports the conclusion of A13 that objects of the same age and SpT may have different spectral signatures of youth.

\begin{table}
\caption{Classification of Pleiades Brown Dwarfs$^a$}
\label{pleiades}
%\begin{center}
\begin{tabular}{lcrl}
\hline
\\
Object & SpT$^b$ & Scores$^c$ & Gravity\\
\hline
\\
PPl~1 & M7    & ?n20 & {\sc int-g} \\
Calar~3 & M8 & 1n22 & {\sc vl-g} \\
Teide~1 & M7 & 2n20 & {\sc vl-g} \\
BRB~17 & L1   & 222n & {\sc vl-g} \\
PLIZ~28 & L0   & 2??2 & {\sc vl-g} \\
PLIZ~35 & L1   & ?22? & {\sc vl-g} \\
BRB~21 & L3    & 2n20 & {\sc vl-g} \\
BRB~23 & L4    & 22?? & {\sc vl-g} \\
\hline
\end{tabular}

$^a$All spectra from \citet{bihain10}. \\
$^b$SpT determined using the method described in A13. \\
$^c$Gravity Scores are listed in the following order: FeH, VO, alkali lines, $H$-band continuum shape.  See A13 for details.
%\end{center}
\end{table}

\section{Atmospheric Models}

\begin{figure}[]
\resizebox{\hsize}{!}{\includegraphics[clip=true]{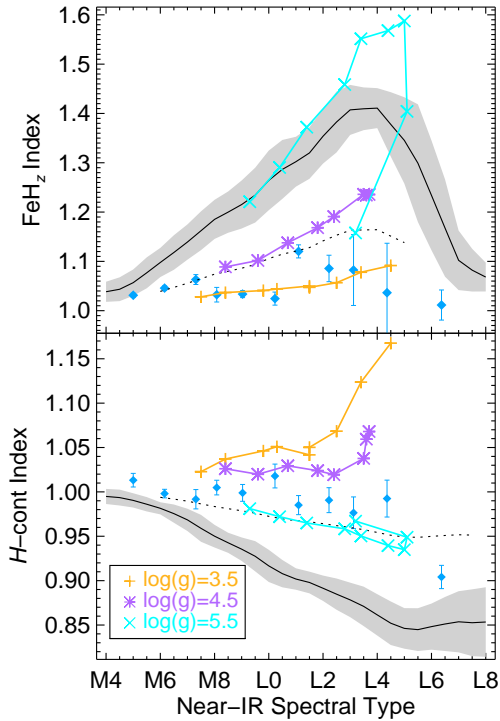}}
\caption{
\footnotesize
Index calculations for BT-Settl model atmospheres.  The models used have T$_{eff}$ of 1800--2600~K in steps of 100K.  For comparison, the {\sc vl-g} standards of A13 are displayed as diamond points.
}
\label{btsettl}
\end{figure}

Atmospheric models are calculated for various values of log(g), which could allow us to tie our gravity classifications to particular log(g) values.  Figure \ref{btsettl} shows the index values calculated for the BT-Settl (AGSS2009) atmospheric models \citep{allard12}.  To place the models on the diagram, we first smoothed and resampled them to have resolution similar to the prism spectra in the A13 sample.  We then treat the model spectra as if they were the spectra of brown dwarfs, determining SpTs and calculating their gravity sensitive indices using the method described in A13.  

A detailed comparison between our spectra and the BT-Settl models is beyond the scope of this work, but a couple of trends became apparent from our index calculations.  Evolutionary models \citep{chabrier00} predict that log(g)=3.5, 4.5, \& 5.5 corresponds to ages of $\sim$5, 50, \& 5000~Myr  for 1800--2600 K objects.   The model FeH$_z$ index values agree fairly well with observations, as do the KI$_J$ indices.  The $H$-Cont index values of the models are significantly higher than observations of objects of similar predicted surface gravity.  The VO$_z$ index for all of the models lie well below the field dwarfs sequence (gray shaded area in Figure \ref{indices}).  

\section{Conclusions}
In conclusion, we have found that binarity and metallicity are unlikely to affect our gravity classifications of young brown dwarfs.  We note, however, that our near-IR spectral types for low-metallicity objects do not show good agreement with their published optical spectral types.  We have applied the A13 classification method to spectra of Pleiades objects from \citet{bihain10}, and find that we classify all of the spectra as having low-gravity, with most being classified as {\sc vl-g}.  A comparison of indices calculated from the BT-Settl model atmospheres shows that the models reproduce the observed FeH$_z$ and KI$_J$ index values reasonably well.  Model VO$_z$ index values, however, are much lower than observations, and model $H$-Cont indices are higher than observations.
 
\begin{acknowledgements}
We are grateful to the organizers of the Brown Dwarfs Come of Age meeting for giving us the opportunity to present our work.  We also thank the participants of the meeting for their helpful and thought-provoking comments, which motivated the discussion presented in this manuscript.  This research has benefited from the M, L, and T dwarf compendium housed at DwarfArchives.org and maintained by Chris Gelino, Davy Kirkpatrick, and Adam Burgasser as well as from the SpeX Prism Spectral Libraries, maintained by Adam Burgasser at http://www.browndwarfs.org/spexprism.   This work was supported by NSF grants AST-0407441 and AST-0507833 as well as NASA Grant NNX07AI83G.

\end{acknowledgements}

\bibliographystyle{aa}

\begin{thebibliography}{}
\bibitem[Allard et al.(2012)]{allard12} Allard, F., Homeier, D., 
\& Freytag, B.\ 2012, Royal Society of London Philosophical Transactions Series A, 370, 2765 

\bibitem[Allers 
\& Liu(2013)]{allers13} Allers, K.~N., \& Liu, M.~C.\ 2013, \apj, 772, 79 

\bibitem[Bihain et 
al.(2010)]{bihain10} Bihain, G., Rebolo, R., Zapatero Osorio, M.~R., B{\'e}jar, V.~J.~S., \& Caballero, J.~A.\ 2010, \aap, 519, A93 

\bibitem[Bowler et al.(2009)]{bowler09} Bowler, B.~P., Liu, 
M.~C., \& Cushing, M.~C.\ 2009, \apj, 706, 1114

\bibitem[Burgasser(2007)]{burgasser07} Burgasser, A.~J.\ 2007, 
\apj, 659, 655 

\bibitem[Burgasser et al.(2010)]{burgasser10} Burgasser, A.~J., 
Cruz, K.~L., Cushing, M., et al.\ 2010, \apj, 710, 1142 

\bibitem[Burgasser 
\& Kirkpatrick(2006)]{burgasser06} Burgasser, A.~J., \& Kirkpatrick, J.~D.\ 2006, \apj, 645, 1485 

\bibitem[Burgasser et al.(2004)]{burgasser04} Burgasser, A.~J., 
McElwain, M.~W., Kirkpatrick, J.~D., et al.\ 2004, \aj, 127, 2856 

\bibitem[Burgasser et al.(2009)]{burgasser09} Burgasser, A.~J., 
Witte, S., Helling, C., et al.\ 2009, \apj, 697, 148

\bibitem[Chabrier et al.(2000)]{chabrier00} Chabrier, G., Baraffe, 
I., Allard, F., \& Hauschildt, P.\ 2000, \apj, 542, 464 

\bibitem[Dahn et al.(2002)]{dahn02} Dahn, C.~C., Harris, 
H.~C., Vrba, F.~J., et al.\ 2002, \aj, 124, 1170

\bibitem[Ducourant et 
al.(2008)]{ducourant08} Ducourant, C., Teixeira, R., Chauvin, G., et al.\ 2008, \aap, 477, L1 

\bibitem[Dupuy 
\& Liu(2012)]{dupuy12} Dupuy, T.~J., \& Liu, M.~C.\ 2012, \apjs, 201, 19 

\bibitem[Faherty et al.(2012)]{faherty12} Faherty, J.~K., 
Burgasser, A.~J., Walter, F.~M., et al.\ 2012, \apj, 752, 56 

\bibitem[Kirkpatrick et al.(2010)]{kirkpatrick10} Kirkpatrick, J.~D., 
Looper, D.~L., Burgasser, A.~J., et al.\ 2010, \apjs, 190, 100

\bibitem[Liu et al.(2013)]{liu13} Liu, M.~C., Dupuy, T.~J., 
\& Allers, K.~N.\ 2013, Astronomische Nachrichten, 334, 85

\bibitem[Looper et al.(2007)]{looper07} Looper, D.~L., 
Burgasser, A.~J., Kirkpatrick, J.~D., 
\& Swift, B.~J.\ 2007, \apjl, 669, L97 

\bibitem[Looper et al.(2008)]{looper08} Looper, D.~L., Kirkpatrick, J.~D., Cutri, R.M., et al.\ 
2008, \apj, 686, 528 

\bibitem[Mann et al.(2013)]{mann13} Mann, A.~W., Brewer, 
J.~M., Gaidos, E., L{\'e}pine, S., \& Hilton, E.~J.\ 2013, \aj, 145, 52 

\bibitem[Weinberger et al.(2013)]{weinberger13} Weinberger, A.~J., 
Anglada-Escud{\'e}, G., \& Boss, A.~P.\ 2013, \apj, 762, 118 

\end{thebibliography}

\end{document}